# Past performance as predictor of successful grant applications

## A case study

**Peter van den Besselaar & Loet Leydesdorff**

**Rathenau Instituut**

Past performance as predictor of successful grant applications

Peter van den Besselaar & Loet Leydesdorff

Preferred citation:
Van den Besselaar, Peter en Loet Leydesdorff, Past performance as predictor of successful grant
applications.  Den Haag, Rathenau Instituut SciSA rapport 0704

**About the authors:**

Peter van den Besselaar is head of the Science System Assessment Department, Rathenau Instituut, and
professor at the Amsterdam School of Communications Research, University of Amsterdam.

Loet Leydesdorff is associate professor at the Amsterdam School of Communications Research, University
of Amsterdam, and visiting professor at the Institute of Scientific and Technical Information of China (ISTIC)
in Beijing, 2007-2010.

# Past performance as predictor of successful grant applications

## a case study

Peter van den Besselaar & Loet Leydesdorff

# Preface

Competitive allocation of research funding is a major mechanism within the science system. It is fundamentally based on the idea of peer review. In fact, the process depends on two different peer review processes: first the selection of papers by journals, which leads to reputation, and second peer review of grant proposals, which is partly based on the reputation and of applicants. Peer review is central in project selection as peers are considered to be in a unique position to identify and select the best and most innovative researchers and research projects. So far the theory, but what can be said about practice?

In this study, we assess the *practice* of peer-review based project selection. The basic question is "do the best researchers get the funding"? In other words, do peer review based quality indicators predict success in receiving funding? The next question to be answered is about the peer review. Do peer reviewers indeed recognize the best researchers, and subsequently, is the decision of the research council based on the outcomes of peer review? Finally, we answer the question whether other factors influence the probability of receiving funding, such as the role of co-applicants, the effects of gender, and the effect of the discipline of the application.

Studying the processes and outcomes of project selection and grant allocation may help to improve the functioning of the research system. We aim to replicate this study in more cases (research fields) and extend it to other aspects of project allocation and funding systems.

This study was conducted in collaboration with the Netherlands Social Science Research council (MaGW/NWO). Without the support of the board and staff of MaGW, this research project would not have been possible. We hope that this study, as well as possible follow-up projects, will contribute to the improvement of research budget allocation and of research policy in general.

Peter van den Besselaar

# Table of contents

# 1 Introduction

Creative researchers are the sources of new ideas. Researchers translate ideas into project proposals, and try to get their projects funded. Research councils develop operating procedures to select the best and most innovative research proposals. In this report we address the question of whether or not this process of variation (by the researchers) and selection (by the research councils) is working in this way. In other words, are research councils selecting the best proposals of the best researchers for funding? The council for social scientific research (MaGW) of the Netherlands Organization for Scientific Research (NWO) kindly provided us with access to their materials for investigating this question in their case.

MaGW/NWO is the research council for the social and behavioral sciences in the Netherlands. It covers all social and behavioral disciplines, including law and economics. It distributes research funds over researchers, institutes and infrastructures, and increasingly plays a leading role in agenda setting, coordination and network formation in its domain. MaGW has a budget of about 36 M€/year, of which is 8 M€ for the open competition, 10 M€ for career grants, and 13 M€ for thematic research grants. In this case study we assess the outcome of the open competition for research funding as well as the career grants, which constitute about half of the councils' budget.

A significant amount of research has been done to investigate how *peer review* systems function, for example, in the context of journals selecting papers for publishing and in the context of grant allocation. Researchers have found that reviewers are not consistent in their assessments of proposals and papers (e.g., Rothwell & Martyn 2000) and some have observed a rather strong gender bias and nepotism (Wenneras & Wold), although this has not been found in more recent periods (Sandstrom & Halsten 2005). A recent meta-analysis confirms the existence of a gender bias, but it does not seem very strong (Bornmann et al, forthcoming).

Peer review comes in a variety of formats, and what is called peer review regularly is actually *committee review*, in which strategic behavior prevails. One of the issues is that committees do no necessarily always select the best amongst their peers, but reach a compromise, allowing each committee member to have his/her favorite proposal funded (Langfeldt 2004).

Another issue is the assessment of interdisciplinary research and research proposals. Research indicates that the assessment of interdisciplinary work is rather problematic, and that this type of work systematically gets low grades (Laudel et al 2006).

This criticism of peer review has resulted in proposals to use *bibliometric quality indicators* as a more objective alternative. However, this has also been criticized, mainly because in some fields bibliometric indicators are not applicable, and because existing bibliographic databases are incomplete and biased, for instance, towards English language publications. Additionally, citing patterns are different between different (sub) disciplines; therefore, bibliometrics indicators are difficult to compare between research fields.

Criticisms concerning peer review have resulted in a variety of proposals to change procedures (Frolich 2003; British Academy 2007); however, some studies do suggest that peer review works well (Wessely 1998; ESF 2006).

This report is organized as follows.

- In chapter 2 the two models used in this study are introduced. Chapter 3 defines the concepts, describes the data and methods used, and ends with the research questions.

- Chapter 4 focuses on the question of whether past performance is related to success in grant applications.

- After having done this, chapter 5 extends the analysis taking into account the referee scores (section 5.1), the discipline to which a proposal belongs (section 5.2), the funding instruments (section 5.3), gender of the applicant (section 5.4), the institutional affiliation of the applicants (section 5.5) and the influence of the co-applicants (section 5.6). Finally, section 5.7 briefly discusses past performance of social and behavioral researchers who publish in scholarly journals but have not applied for a research grant, and compares this with the past performance of the applicants.

- The report ends with the conclusions and with a discussion of the implications and open issues (chapter 6).

# 2    The model

This study's basic question will be answered in terms of the relationship between the past performance of researchers and the allocation decision of a research council: is research money distributed among the best researchers? Figure 1 shows the basic model, in which we consider the research council's procedure as a black box. We are interested in the relation between the input and the output.

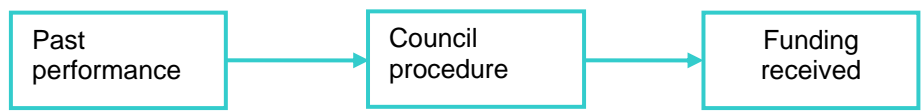

**FIGURE 1:** Basic model

The main issue is to define past performance. In this study we use the number of publications and number of citations received by the applicant to measure his or her past performance, as well as the quality of the co-applicant(s). Initially, the focus was on past performance in terms of the number of publications produced by an applicant in the two years preceding the application, plus the year of application, but it turned out that in some cases the number of citations during the same period was the better predictor (despite this short citation window).

Since this is a case study about a social science research council, the *Social Science Citation Index* (*SocSCI*) is used as a comprehensive data source, even though this has been disputed within the social sciences (SWR/RGW 2005).  A few issues are relevant to our study. First, in several subfields of the social sciences, books and not journals are the main publication format. We do not take these other publication formats into account, and for several fields (such as law) this may be a serious disadvantage. For those fields, the current analysis should be interpreted with even more care (Nederhof 2006). Second, the journals indexed in the *SocSCI* do not cover the total relevant journal space, and the coverage of the *SocSCI* is different for different parts of the social sciences. Elsewhere we have shown that the orientation on *SocSCI* processed journals differs per subfield (Van den Besselaar 2007). Third, even if sub-fields of the social sciences are equally well covered by the *SocSCI*, they are not homogenous in terms of publication patterns and citation patterns.

Increasingly, Dutch universities are operationalizing quality in terms of citations and publications in ISI-indexed journals, for example, in the evaluation of graduate schools. Although many researchers have doubts about the validity of the (*Soc)SCI*'s counting of publications and citations as an indicator for scientific quality, the funding and evaluation procedures increasingly take into account this type of indicator; therefore researchers themselves try to get into the 'top journals' as defined by the *SocSCI*'s impact factor (Leydesdorff, *Jasist,* in press).

After the test of our basic model, we will extend it to a few other factors. Here, we will not discuss the theoretical background in depth, but only describe it briefly. Figure 2 shows the extended model.

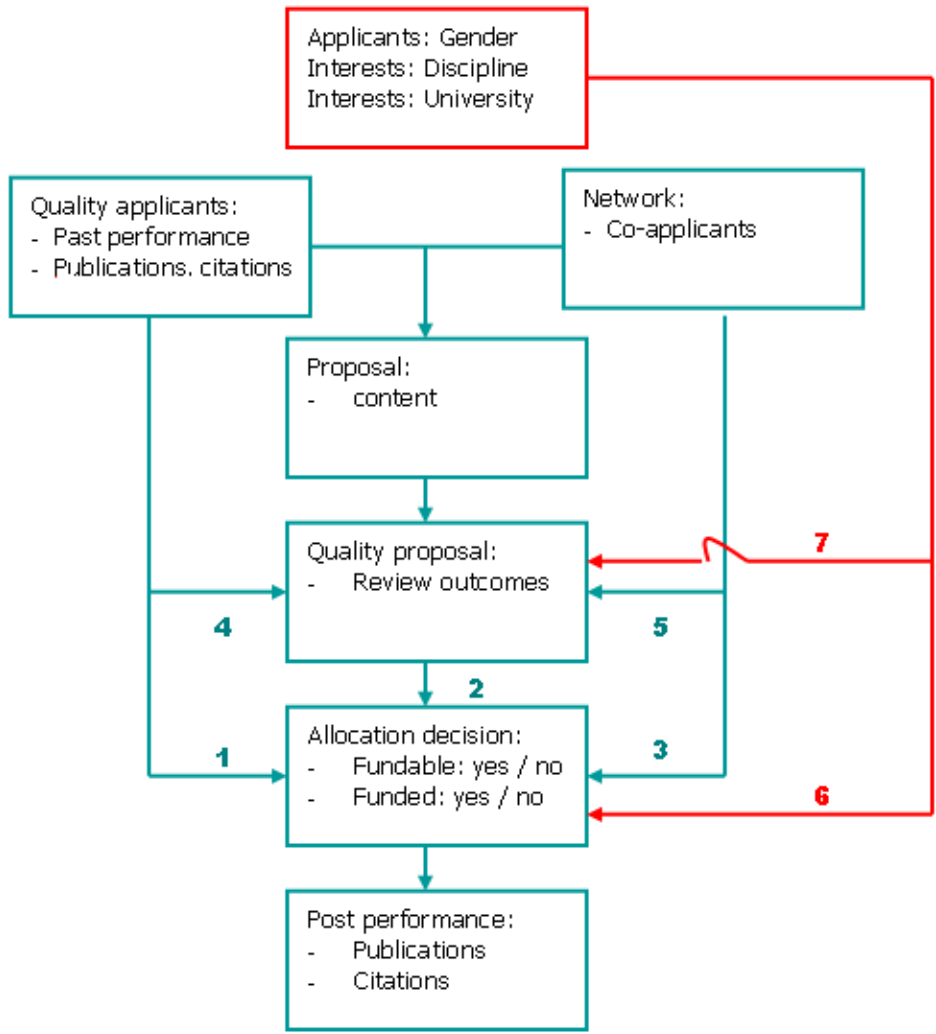

**FIGURE 2:** Extended model

The extended model is based on a combination of a Mertonian and a constructivist perspective on peer review (Bornmann 2007). The constructivist perspective adds that the social structure has to be reproduced by agency and that choices have to be made. Decisions made about the funding of research, for example, are based on the quality of the researcher involved (1), the proposal (2), and the network of the applicant (3). The quality of the proposal is defined by the judgments of peers, which are based on a review process. In this judgment, the overall quality of the applicant (4) and his/her network (5) may play a role.[1] One should not easily assume that the Mertonian norms of science (Merton, 1942) will prevail in the selection process (Wenneras & Wold, 1997).

By adding a 'social constructivist' perspective to the Mertonian scheme, one can also take into account certain interests and social factors. In this study, we include three variables that reflect these factors. We will also test whether gender, discipline and university affiliation differences are related to the decisions of the research council (6), and to the referees' judgments (7).

Of course, the past performance of the researcher and his/her network can also influence the quality of the proposal; thus reviewers may also take into account past performance, rather than just the quality of the specific application under review. In this study we restrict ourselves to the relationship between past performance and the decision made about the application (arrow 1); the effect of the reviewers' assessments (arrow 2), and the quality of the applicant's network (arrow 3).

We also do some tests about the effects of past performance (arrow 4) and the applicants' network (arrow 5) on the review. We also focus on the following question: how do contextual factors play a role? This question is important, because it may teach us something about the quality of the procedures: are they biased in one way or another? The relationship between the 'mertonian variables' and the outcome of the decision tend to reflect little bias and convey more about the effectiveness of the selection process. To measure the latter more directly would require the inclusion of post-performance. However, for the dataset under analysis here (applications to the research council in 2003, 2004 and 2005), this is too early.

---

[1] However, in case of the young researchers program, not so much past performance but expected future performance may be an importance consideration.

# 3    Data and methods

In this study the independent variable is the decision about the proposal:
- 'fundable' and funded applications (A)
- 'fundable' but not funded applications (A-)
- 'non-fundable' applications (B)[2]

*Groups of researchers:* Since researchers do not submit an application every year, three different groups are distinguished in parts of this analysis:
- Successful applicants
- Unsuccessful applicants
- Non-applicants[3]

*Disciplines and years:* Citation behavior and orientation on SSCI journals differ per subfield; thus it may be useful to distinguish between subfields. In this report, we start at the aggregate level of the social and behavioral sciences as a whole, and cover the three years for which we were provided with data: 2003, 2004, and 2005. After that we check whether the results would be different if the analysis was done for the years separately, and for subfields separately.

*Instruments:* The research council uses a variety of funding instruments, and the data distinguish between personal grants for researchers in different phases of their careers, the 'vernieuwingsimpuls' (VI) and the open competition (OC).[4] The role of past performance differs per instrument, so there is also a need for distinguishing between instruments. Specifically, the personal grants for starting researchers (the VENI awards) may not be based on past performance, since these researchers are probably too young to have a strong track record. As a result, we also analyzed the data for the four instruments separately.

*Past performance:* We define the quality of the applicant in terms of past performance. And we operationalize past performance in a specific way. For each of the years, the time horizon for past performance is three years. In other words, success in 2003 is related to performance data from 2001 to 2003, success in 2004 to performance data from 2002 to 2004, and success in 2005 to performance data from 2003 to 2005.

---

[2] If we group A- and A together, in stead of A- an B, the results of the analysis do not change in a meaningful way.

[3] If an allocation mechanism would attract the best researchers, the non-applicants are expected to score lower on the quality criteria than the applicants.

[4] The thematic programs of the research council cannot be included in this analysis. It would be useful to analyze these too, as the goals of the thematic programs are different – and include criteria such as the possible societal outcomes of the research.

The "times cited" are measured on Feb. 9, 2007. All papers (articles, reviews, and letters in the three year periods) with a Dutch address are included. The applications to MaGW are the units of analysis; in the case of non-applicants the author names are used. In other words, when a single researcher applies more than once in one year, s/he may score differently on the grants received, but not on the quality indicators. If a researcher applies in different years, past performance is different between the years, as publications and citations for the same applicant may differ between the periods (e.g., 2001-2003; 2002-2004; 2003-2005).

The last name and the first initial were used for matching the ISI data and the application data. This generates a bit of error. For example, one of Leydesdorff's articles is listed in the ISI-database as authored by "T. Leydesdorff", and Van den Besselaar also appears as "Van den BesselaarA" in the database. These errors were not (manually) corrected. Our results should therefore be read as statistics with margins of error.

*Quality of the network*: In this study we define the quality of the applicants' network as the quality if the co-applicants. The past performance of the co-applicants is operationalized in the same way as past performance of the applicants - that is using publications and citations. More specifically, we use two different operationalizations to obtain indicators for the quality of the network:

- Average network quality
  - average number of publications by the applicant and co-applicants
  - average number of citations received by the applicant and co-applicants
- Maximal network quality
  - number of publications of the most publishing of the applicant and the co-applicants
  - number of citations received by the most cited of the applicant and the co-applicants

Of course, this has two implications. First, only open competition applications can be included in the analysis, as the other three programs do not require co-applicants. And second, other dimensions of an applicants' network are neglected, such as the status of the applicants' PhD supervisor, the co-author network of the applicant, and the like. This is an interesting topic for further research.

*Quality of the proposal*: The database includes the referee's judgments. However, different instruments have a different classification system, which also changes over the years. Three of the four scales are a five-point scale, the other a three-point scale. We translated these into one five-point scale, as shown in table 1.

*Methods:* four methods will be used in the analysis:

- Correlation analysis: the relation between the independent variables (past performance, quality of the application, quality of the network, referee's judgment) and the dependent variable (the awarded grants)?
- Analysis of variance: are successful applicants different from the non-successful in terms of the independent variables?
- Discriminant analysis: can we 'predict' which application will be successful, based on the independent variables.
- Visual inspection of the distributions, in order to better interpret the statistical analysis.

Using correlation analysis, we relate past performance of the applicant and the review score of the application (the independent variables) with the amount of money received. Using Anova, we compare the group of successful applications with the group of unsuccessful applications: do they differ in terms of the independent variables. Using Discriminant Analysis, we test if the independent variables can be used to 'predict' whether an application is successful or not. In the latter two analyses, the amount of money received is not taken into account.

**TABLE 1:** Coding of the referees' scores

| Coding | OC | OC / VI | VI* |
|--------|----|---------|-----|
| 1 | A | Excellent | Continue |
| 2 |  | Very Good | Doubt/continue |
| 3 | B | Good | Doubt |
| 4 |  | Fair | Doubt/stop |
| 5 | C | Poor | Stop |

OC: open competition
VI: 'vernieuwingsimpuls'
*    used in the preselection phase

*Research questions*: We will answer the following questions in this report:
1. Are the A (funded), A- (fundable, unfunded) and B (unfunded) applicants different in terms of past performance?
2. Does past performance (of all the three groups) correlate with funding received?
3. Can we predict the success of applicants from their past performance?

Then we turn to the elaborated model, and include the referees' judgments:
4. Are the A (funded), A- (fundable, unfunded) and B (unfunded) applicants different in terms of number of
5. Can we predict the success of applicants based on their numbers of publications, citations, and referee scores (and in case of the open competition: the quality of the co-applicants)?

After answering these central questions, we then analyze the influence of some mediating variables, such as different subfields, funding instruments, and gender. More specifically, we will discuss the following four issues:

6. Are there differences between disciplines: are the results different in e.g., law, economics, psychology and other disciplines under study here? Does the discipline influence the relations between the performance variables and the probability of success?

7. Are there differences between the instruments: the 'open competition' and the three 'career development programs' (Veni, Vidi, Vici)?

8. And what about gender differences?

9. Differences between universities? is the university affiliation of the main applicant related to the probability of success?

10. Have patterns changed over the years?

11. What about those who did not apply? Do the successful applicants (A) differ from the unsuccessful (A- and B) applicants and from the non-applicants?

**TABLE 2:** Variables in this study

| | |
|---|---|
| Pub | Number of publications by the applicant (three years before the application) |
| Cit | Number of citations to these publications at 7 Feb. 2007 |
| Pub2 | Average number of publications by applicant and the co-applicants |
| Cit3 | Average number of citations to the publications of the applicant and the co-applicants, at 7 Feb. 2007 |
| Pub3 | Number of publications by the most productive of the applicant and the co-applicants |
| Cit3 | Number of citations to the publications of the most cited of the applicant and the co-applicants, at 7 Feb. 2007 |
| Sex | Gender of main applicant |
| Uni | University of main applicant |
| Disc | Discipline of application |
| Instr | funding instruments (three types of personal grants and open competition) |
| Ref | Average of the referee's reports |
| Dec | 'Fundability': assessment by research council |
| euro | 'Funding': grants received from the research council in (euro). |

# 4 The basic model

## 4.1 Past performance and budget allocation

ANOVA can be used to test whether recipients score higher in terms of output and citations than researchers who did not get their proposal accepted. Table 3 gives the results. The two groups score differently in terms of the average numbers of publications and received citations. The successful applicants publish more than the failed applicants (4.45 versus 2.71 publications) and are cited more (36 versus 15 citations). All differences are statistically significant. Finally, the referees are significantly more positive about the funded applications. The difference is one point in a five point scale. The successful applications score 1.6 ('very good/excellent') and the non-funded applications score on average 2.7 (slightly better than 'good')

**TABLE 3:** Publications and citations by success (2003-2005)

|  |  | N | Mean | Std. Deviation | Std. Error | 95% Confidence Interval for Mean Lower Bound | 95% Confidence Interval for Mean Upper Bound | Minimum | Maximum |
|---|---|---|---|---|---|---|---|---|---|
| Citations | A | 275 | 36.03 | 70.704 | 4.264 | 27.64 | 44.43 | 0 | 593 |
|  | A- / B | 911 | 15.61 | 45.295 | 1.501 | 12.67 | 18.56 | 0 | 621 |
|  | Total | 1186 | 20.35 | 52.969 | 1.538 | 17.33 | 23.36 | 0 | 621 |
| Publications | A | 275 | 4.45 | 5.988 | .361 | 3.74 | 5.16 | 0 | 43 |
|  | A- / B | 911 | 2.71 | 4.915 | .163 | 2.39 | 3.03 | 0 | 62 |
|  | Total | 1186 | 3.11 | 5.233 | .152 | 2.81 | 3.41 | 0 | 62 |

|  |  | Sum of Squares | df | Mean Square | F | Sig. |
|---|---|---|---|---|---|---|
| Citations | Between Groups | 88091.424 | 1 | 88091.424 | 32.224 | .000 |
|  | Within Groups | 3236713.147 | 1184 | 2733.710 |  |  |
|  | Total | 3324804.571 | 1185 |  |  |  |
| Publications | Between Groups | 638.992 | 1 | 638.992 | 23.785 | .000 |
|  | Within Groups | 31808.317 | 1184 | 26.865 |  |  |
|  | Total | 1332.180 | 1177 |  |  |  |

A: funded ; A-:fundable, not funded; B: not-fundable

What changes if we group the applications in different way: the 'fundable applications' (A and A-) versus the 'non-fundable applications' (B)? Again, the 'fundable' applications score significantly better in all variables than the 'non-fundable' ones. The differences are statistically significant but at the same time smaller than in the comparison of the funded (A) and the non-funded (A- and B) applications. Indeed, comparing A with A-, again the A's score significantly higher than the A-'s applications.[5]

Most statistical techniques used in this study are meant for data with normal distributions, although Discriminant Analysis is said to be robust against violation of the assumptions. Because the data used have a rather skewed distribution, we also analyze the data visually. This helps to interpret the statistical results correctly. Let us first present graphs of the distribution of publications and of citations.

---

[5] Table 3.1 and 3.2 in the appendix give the details.

Both figures show the funded applications on the left side and the rejected ones on the right side. The thin (in color: red) line represents the citations and the thick (blue) line the publication of the applicant. Clearly, the distributions of both variables are skewed – as expected. Figure 3 suggests that the distribution of the number publications per applicant is not very different between the two groups. Figure 4 suggests the same for the distribution of the citations. In any case, if there are differences, they are not large.

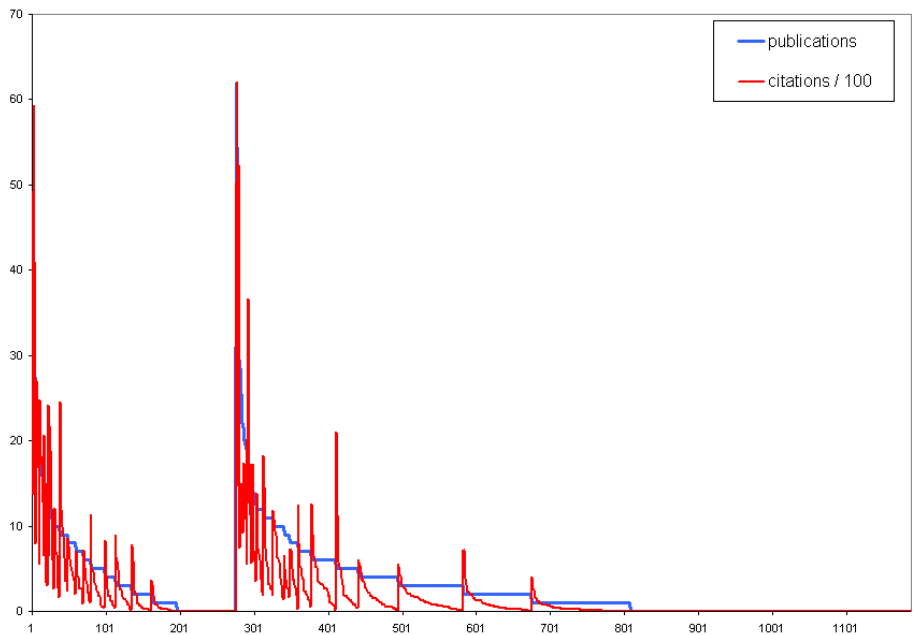

**FIGURE 3:** Distribution of publications; successful applicants (left) and unsuccessful (right)
(left axis: number of publications and of citations/100)

To obtain a more detailed overview, we also plot the distribution of publications of the 275 successful applicants and the top performing 275 non-successful applicants in one graph (figure 5).[6] The same was done for the citations (figure 6). The results are interesting: the best of the unsuccessful applicants actually score higher in terms of past performance than the successful applicants do. In other words, the long tail of applicants with low past performance causes the differences in the averages between the groups. Therefore, we can conclude that the skewed distribution of the data indeed influences the statistical results.

---

[6] Based on citations received and (secondly) on publications.

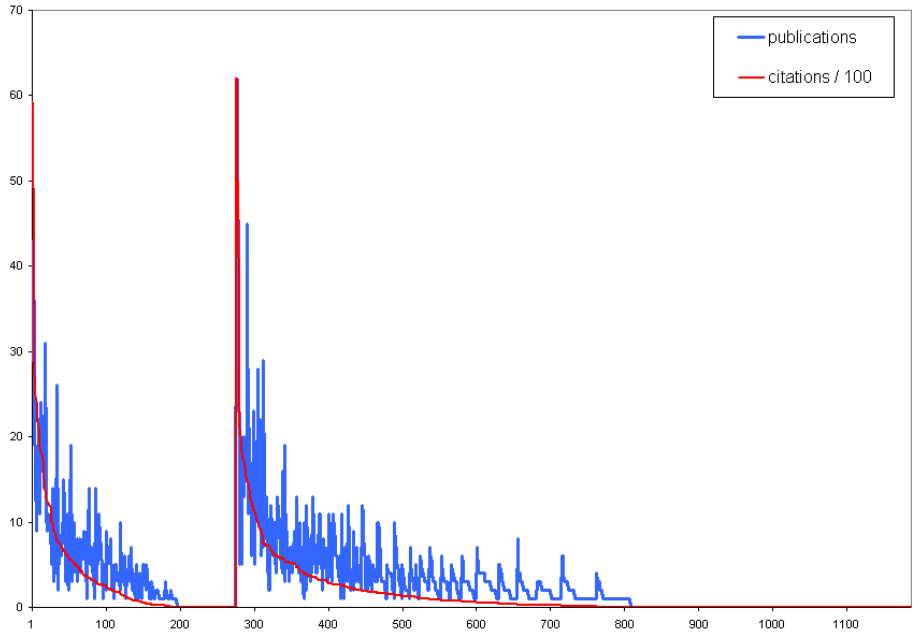

**FIGURE 4:** Distribution of citations; successful applicants (left) and unsuccessful (right)
(left axis: number of publications and of citations/100)

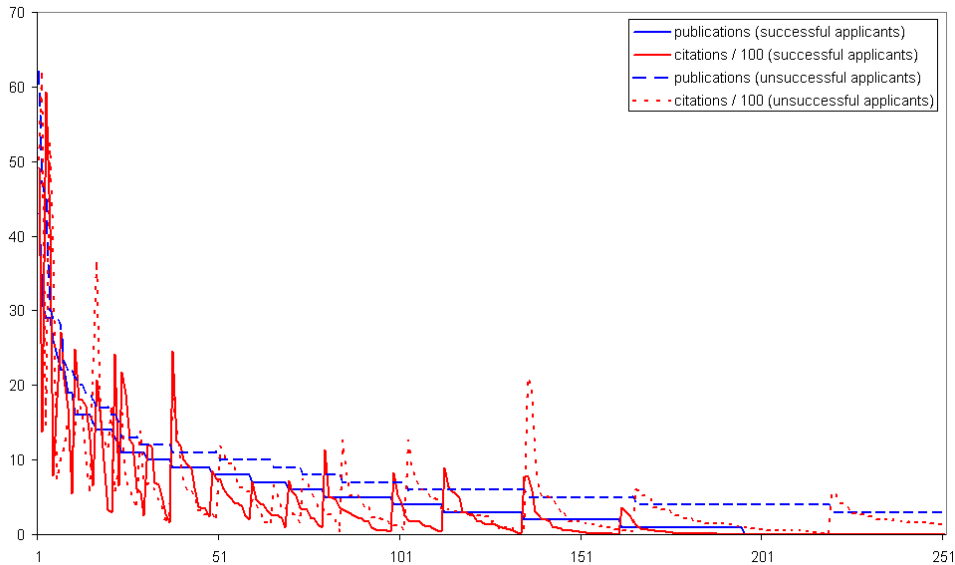

**FIGURE 5:** Distribution of publications of successful and best unsuccessful applicants
(left axis: number of publications and of citations/100)

Restricting the ANOVA to the top 275 unsuccessful applicants and the 275 successful one's radically changes the result. As already suggested by figures 5 and 6, unsuccessful applicants score significantly better in both past performance indicators. The distribution over the four instruments (OC, Veni, Vidi, Vici) is not equal in both groups.[7] Correcting for this – by using a stratified sample of the best unsuccessful applicants – does not change this result. Table 4 shows the results of the ANOVA for the stratified sample of the top unsuccessful applicants.

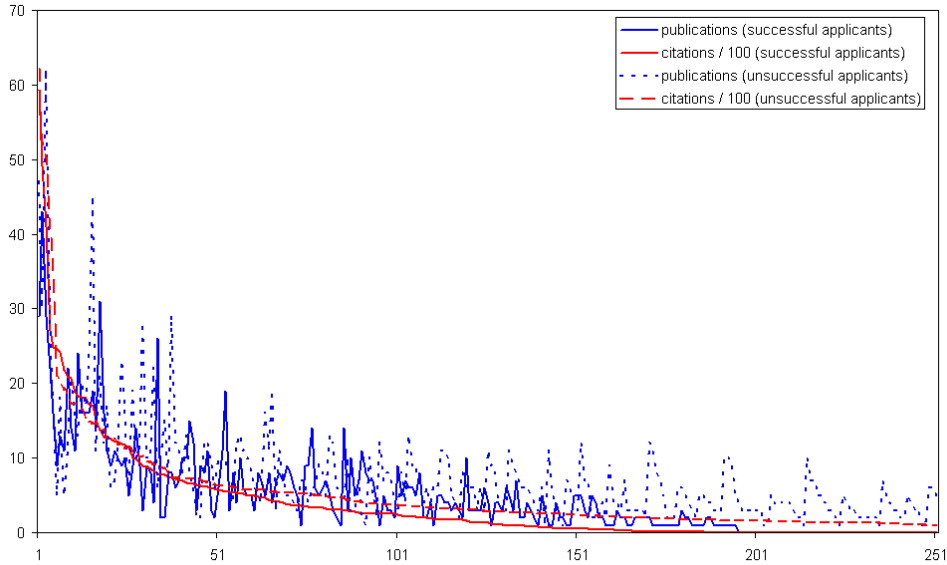

**FIGURE 6:**
Distribution of citations of successful and best unsuccessful applicants
(left axis: number of publications and of citations/100)

**TABLE 4:** Publications and citations by success (2003-2005)

|  |  | N | Mean | Std. Deviation | Std. Error | 95% Confidence Interval for Mean | | Min | Max |
|---|---|---|---|---|---|---|---|---|---|
|  |  |  |  |  |  | Lower Bound | Upper Bound |  |  |
| pub | A | 275 | 4.44 | 5.992 | .361 | 3.73 | 5.15 | 0 | 43 |
|  | A-/B* | 277 | 6.92 | 7.068 | .425 | 6.08 | 7.75 | 0 | 62 |
|  | Total | 552 | 5.68 | 6.664 | .284 | 5.13 | 6.24 | 0 | 62 |
| cit | A | 275 | 36.03 | 70.708 | 4.264 | 27.63 | 44.42 | 0 | 593 |
|  | A-/B* | 277 | 48.04 | 72.338 | 4.346 | 39.48 | 56.59 | 0 | 621 |
|  | Total | 552 | 42.05 | 71.718 | 3.053 | 36.06 | 48.05 | 0 | 621 |

|  |  | Sum of Squares | Df | Mean Square | F | Sig. |
|---|---|---|---|---|---|---|
| pub | Between Groups | 846.670 | 1 | 846.670 | 19.711 | .000 |
|  | Within Groups | 23624.850 | 550 | 42.954 |  |  |
|  | Total | 24471.520 | 551 |  |  |  |
| cit | Between Groups | 19907.016 | 1 | 19907.016 | 3.891 | .049 |
|  | Within Groups | 2814158.461 | 550 | 5116.652 |  |  |
|  | Total | 2834065.476 | 551 |  |  |  |

A: funded ; A-:fundable, not funded; B: not-fundable
*: stratified (by funding instrument) sample of best scoring unsuccessful applicants

With these results, question 1 can be answered. On average:
- the funded applicants have a better past performance than the non-funded;

---
[7]  The distribution in the successful group: OC=154; Veni=65; Vidi=43; Vici=12.
   The distribution in the unsuccessful top: OC=159; Veni=83; Vidi=21; Vici=12.

- the fundable applicants have a better past performance than the non-fundable;
- the funded applicants have a better past performance than the fundable-non-funded;
- however, the 'best' non-funded applicants have a significantly better past performance than the funded applicants. This suggests firstly that selection is not strongly based on past performance (as defined in this study), and secondly that the reservoir of potential recipients is much larger than the group that was funded (Melin and Danell 2006).[8]

## 4.2     Can we predict success from past performance?

We will now analyze the relationship between past performance and success in getting an application funded. In table 5, we give the correlation between the independent variables and the amount of funds received from the research council. The application is again the unit of analysis. Consequently, if researchers file more than one project, these projects are treated as two different cases.

**TABLE 5.**  Success by past performance

|  |  | Cit | euro |
|---|---|---|---|
| Publications | Pearson Correlation | .818(**) | .159(**) |
|  | Sig. (2-tailed) | .000 | .000 |
|  | N | 1186 | 1186 |
| Citations | Pearson Correlation |  | .197(**) |
|  | Sig. (2-tailed) |  | .000 |
|  | N |  | 1178 |

As table 5 shows, past performance measures – numbers of citations received and number of publications – correlate strongly. They correlate also with the amount of funding received, but this correlation is low. Because the data used are relatively skewed (publications, citations), an ordinal measure of association such as spearman's rho is preferred.[9] The same pattern emerges as in the case of the Pearson's correlation, but the correlations are stronger (table 6).

---

[8] A recent evaluation of the 'VI programs' of the research councils Geosciences and Life Sciences (Aard- en Levenswetenschappen,), Chemical Sciences & Advanced Chemical Technologies for Sustainability), and Physical Sciences (Exacte Wetenschappen) shows similar results (Van Leeuwen 2007). Using a different indicator for past performance, Van Leeuwen shows that in most cases average past performance of the awarded researchers is higher than the non-awarded one's. However, he did not compare between the successful applicants with the best non-successful ones. Our hypothesis would be that the relations found for these research councils disappear if the analysis would be done for the best applicants only.

[9] The referees' judgments are less skewed, and the mean (2.42) and median (2.17) only differ slightly.

**TABLE 6.** Success by past performance

|  |  | cit | euro |
|---|---|---|---|
| Publications | Spearman's rho | .923(**) | .160(**) |
|  | Sig. (2-tailed) | .000 | .000 |
|  | N | 1186 | 1186 |
| Citations | Spearman's rho |  | .185(**) |
|  | Sig. (2-tailed) |  | .000 |
|  | N |  | 1178 |

** Correlation is significant at the 0.01 level (2-tailed)

Can one predict success using the independent variables? Again, for this we use a Discriminant Analysis (DA). DA may be used specifically to predict group membership (a nominal variable) from interval variables, such as (here) the numbers of publications and citations.

Firstly, we conduct a DA, using only the number of publications and citations received by the applicant. Table 7 shows the results. Based on past performance, about one third of the funded projects are classified correctly, and about 85% of the rejected applications. Actually, running the stepwise procedure, about the same result occurs. But the DA only includes the received citations only (table 8). All the discriminant analyses de result in a significant model.

**TABLE 7:** Success of applications by pub en cit

|  |  | A versus A-/B | Predicted Group Membership | | Total |
|---|---|---|---|---|---|
|  |  |  | A | A- or B |  |
| Original | Count | A | 99 | 176 | 275 |
|  |  | A- or B | 151 | 760 | 911 |
|  | % | A | 36.0 | 64.0 | 100.0 |
|  |  | A- or B | 16.6 | 83.4 | 100.0 |

72.4% of original grouped cases correctly classified.
A = funded; A- = fundable, not funded; B = not fundable

**TABLE 8:** Success of applications by pub en cit (stepwise*)

|  |  | A versus A-/B | Predicted Group Membership | | Total |
|---|---|---|---|---|---|
|  |  |  | A | A- or B |  |
| Original | Count | A | 96 | 179 | 275 |
|  |  | A- or  B | 142 | 769 | 911 |
|  | % | A | 34.9 | 65.1 | 100.0 |
|  |  | A- or B | 15.6 | 84.4 | 100.0 |

72.9% of original grouped cases correctly classified.
* Stepwise: *cit* in the analysis

In summary, successful and unsuccessful applications differ in terms of the numbers of publications by the applicants, but these differences do not differentiate between the two classes. The model correctly classifies some 35% of successful applications, and some 85% of the unsuccessful. This does not change in the stepwise model. In other words, on this level of all instruments and all sub-disciplines, the past performance (as defined in this specific way) is only weakly related to success.

**TABLE 9.** Success by past performance*

|       |                   | Cit       | euro      |
|-------|-------------------|-----------|-----------|
| Pub   | Spearman's rho    | .833(**)  | -.256(**) |
|       | Sig. (2-tailed)   | .000      | .000      |
|       | N                 | 552       | 552       |
| Cit   | Spearman's rho    |           | -.262(**) |
|       | Sig. (2-tailed)   |           | .000      |
|       | N                 |           | 552       |

\*    Stratified (by funding instrument) sample of best scoring unsuccessful applicants
\*\*  Correlation is significant at the 0.01 level (2-tailed).

As in section 4, the long tail associated with the unsuccessful group may strongly influence the outcome of the analysis. We therefore repeated the analysis including only the top of the unsuccessful applicants (stratified sample in terms of funding instruments). The results are striking: the correlation between past performance and funding gets negative, and the correlations are stronger than in case of the whole sample (table 9).

The Discriminant Analysis works slightly different than it does in the case of the whole sample. A larger number of the successful applications is correctly classified now. However, the number of correctly classified unsuccessful application has decreased (table 10).

**TABLE 10.** Classification of applications from pub en cit (stepwise*)

|          |       | decision | Predicted Group Membership | | Total |
|----------|-------|----------|------|---------|-------|
|          |       |          | A    | A- or B |       |
| Original | Count | A        | 192  | 83      | 275   |
|          |       | A- or B  | 148  | 129     | 277   |
|          | %     | A        | 69.8 | 30.2    | 100.0 |
|          |       | A- of B  | 53.4 | 46.6    | 100.0 |

a  58.2% of original grouped cases correctly classified.
A = funded; A- = fundable, not funded; B = not funded

We can now answer the second and the third question: Do past performance and success correlate? Can we predict the success of applicants from their publications and citations?

- The correlation analysis shows only a low correlation between the scientometric indicators of past performance, and the amount of money received.
- If we want to predict success and failure (and do not take into account the amount of money received), the number of citations can be used to predict group membership. However, only a low percentage of the successful cases are correctly classified.
- Finally, within the group of successful and top 275 unsuccessful applicants, the relationship is completely opposite: we find a negative correlation between past performance and the amount of funding received, and between referee judgment and funding.

# 5 The extended model

## 5.1 Referee scores and success

Now we extend the model, especially with the scores of the reviewers. ANOVA can be used to test whether the recipients score higher in terms of referee judgments than the researchers that did not get their proposal accepted (table 11)? The referees were significantly more positive about the funded applications. The difference is one point in a five point scale. The funded applications score 1.6 ('very good/excellent') and the non-funded applications score on average 2.7 (slightly better than 'good').

What changes if we group the applications in different way: the 'fundable applications' (A and A-) versus the 'non-fundable applications' (B)? Again, the 'fundable' applications score significantly better in all variables than the 'non-fundable' ones. The differences are statistically significant but at the same time smaller than in the comparison of the funded (A) and the non-funded (A- and B) applications. Indeed, comparing A with A-, again the A's score significantly higher than the A-'s applications.

**TABLE 11:** Referee results by success (2003-2005)

| | | N | Mean | Std. Deviation | Std. Error | 95% Confidence Interval for Mean Lower Bound | 95% Confidence Interval for Mean Upper Bound | Minimum | Maximum |
|---|---|---|---|---|---|---|---|---|---|
| Referee | A | 274 | 1.5929 | 0.6343 | .03645 | 1.5211 | 1.6647 | 1.00 | 3.67 |
| Judgment | A-/ B | 904 | 2.6770 | 1.04475 | .03475 | 2.6088 | 2.7452 | 1.00 | 5.00 |
| | Total | 1183 | 2.4249 | 1.06388 | .03100 | 2.3640 | 2.4857 | 1.00 | 5.00 |

| | | Sum of Squares | Df | Mean Square | F | Sig. |
|---|---|---|---|---|---|---|
| Referee | Between Groups | 247.143 | 1 | 247.143 | 267.862 | .000 |
| | Within Groups | 1085.037 | 1176 | .923 | | |
| | Total | 1332.180 | 1177 | | | |

A: funded ; A-:fundable, not funded; B: not-fundable

We now proceed with the correlation between referees' scores and success in getting an application funded. Table 12 shows the correlation between the independent variables and the amount of funds received from the research council. Again, the application is the unit of analysis, so if researchers did successfully get more than one project funded, these projects are treated as two different cases.

The referees' judgments correlate low but significantly with the past performance indicators. It also correlates moderately with the funding received, but higher than the past performance indicators do. Please note that the negative sign in this case is actually pointing to a positive correlation. The scale used for measuring the referee's judgment uses 1 for 'very good' and 5 for 'poor'.

**TABLE 12** Success by past performance and peer review

| | | Cit | Referee | euro |
|---|---|---|---|---|
| Publications | Pearson Correlation | .818(**) | -.173(**) | .159(**) |
| | Sig. (2-tailed) | .000 | .000 | .000 |
| | N | 1186 | 1178 | 1186 |
| Citations | Pearson Correlation | | -.179(**) | .197(**) |
| | Sig. (2-tailed) | | .000 | .000 |
| | N | | 1178 | 1186 |
| Referee | Pearson Correlation | | | -.326(**) |
| | Sig. (2-tailed) | | | .000 |
| | N | | | 1178 |

Also here the ordinal measure of association (spearman's rho) gives a similar pattern as in the case of Pearson's correlation, but the correlations are stronger. Especially the correlation between the referee's judgment and the amount of money received is moderately strong now. (table 13)

**TABLE 13:** Success by past performance and peer review

| | | cit | referee | euro |
|---|---|---|---|---|
| Publications | Spearman's rho | .923(**) | -.205(**) | .160(**) |
| | Sig. (2-tailed) | .000 | .000 | .000 |
| | N | 1186 | 1178 | 1186 |
| Citations | Spearman's rho | | -.214(**) | .185(**) |
| | Sig. (2-tailed) | | .000 | .000 |
| | N | | 1178 | 1186 |
| Referee | Spearman's rho | | | -.455(**) |
| | Sig. (2-tailed) | | | .000 |
| | N | | | 1178 |

** Correlation is significant at the 0.01 level (2-tailed)

Can we predict success using the independent variables? Again, we use Discriminant Analysis (DA) and also include the variable *results of the referee process* in the analysis. The outcome of the DA is much better now (table 14 versus table 8), as the percentage correctly classified successful applications increases to 85%. The number of correctly classified unsuccessful ones is now slightly lower at 61%. A stepwise procedure does not change the result – but the 'pub' variable is removed again.

**TABLE 14:** Classification of applications from pub, cit, and ref (all, stepwise*)

| | | A versus A- / B | Predicted Group Membership | | Total |
|---|---|---|---|---|---|
| | | | A | A- or B | |
| Original | Count | A | 231 | 43 | 274 |
| | | A- or B | 354 | 550 | 904 |
| | % | A | 84.3 | 15.7 | 100.0 |
| | | A- or B | 39.2 | 60.8 | 100.0 |

a 66.3% of original grouped cases correctly classified.
* Stepwise: *cit* and *ref* in the analysis

All the discriminant analyses result in a significant model. Can we predict the success of applicants from their publications, citations, and referee scores? The referees' judgments contribute much to the correct classification. Also, the relation between money received and the referee's judgments is moderately high at 0.46. Nevertheless, this implies that still a large part of the variance remains unexplained. Is this related to the dispersion of the reviewers' judgments?

Figures 7, 8, and 9 do suggest this. They plot the variance of the reviews of a proposal against the average score of the referees. Apart from a few high scoring applications, the variance indeed is large in most of the cases.

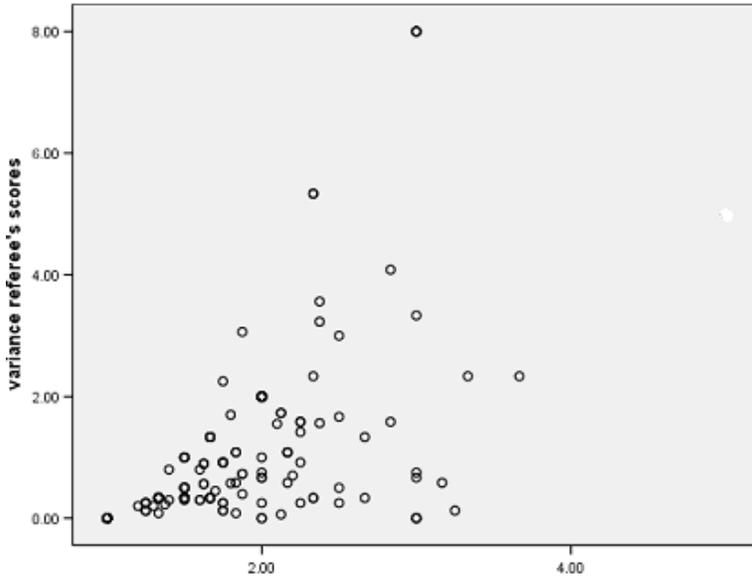

**FIGURE 9:** Variance by average referees' score - the funded proposals (A)

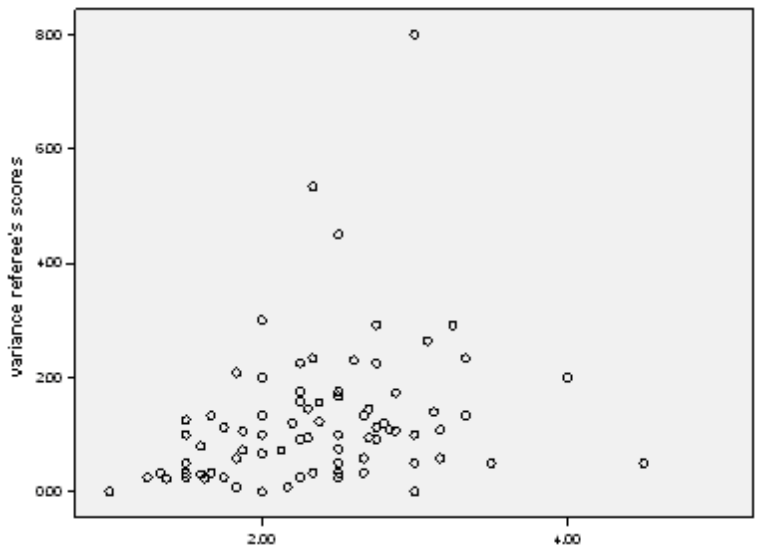

**FIGURE 10:** Variance by average referees' score– fundable non-funded (A-)

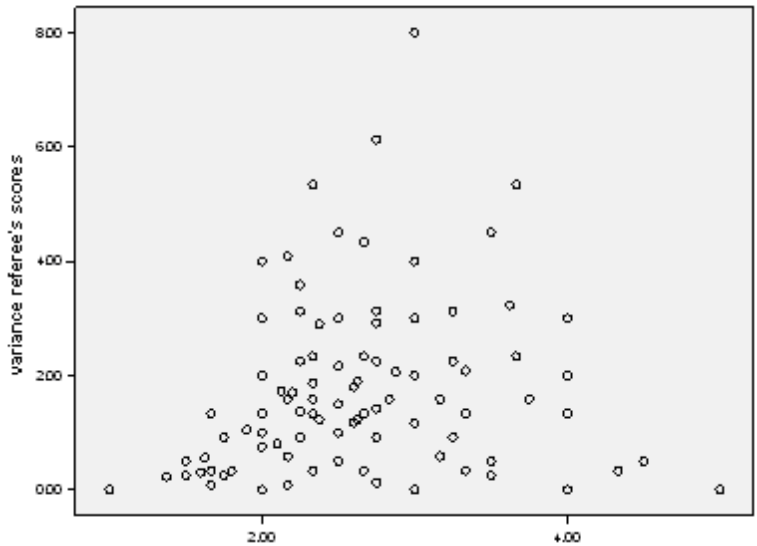

**FIGURE 11:** Variance by average referees' score– not-fundable (B)

Additionally, the scatter plots are rather similar: so we have high scoring (average between 1 and 1.5) applications with low variance (between 0 and 1) in all three categories: the funded (A), fundable non-funded (A-) and non-fundable (B).

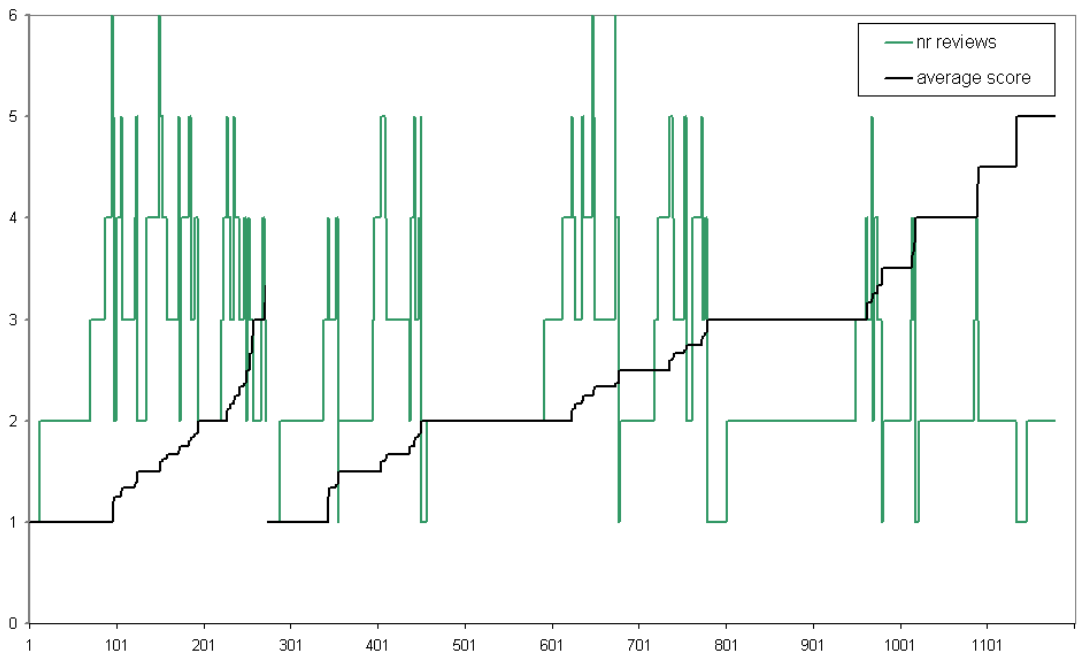

**FIGURE 10:** Distribution of average referees' score (black line); successful applicants (left) and unsuccessful (right). Grey (in color: green) line: number of reviews*

(*: The large variation in the number of reviews is partly due to the different procedures in programs under study. In some cases the procedure includes a pre-selection, resulting in a higher number of reviews)

Next, we compare the scores received from the reviewers visually (in figure 12). The successful applicants seem to do better than the unsuccessful ones. The number of applications receiving a "1" is relatively larger in the left part of the graph than in the right, and the opposite is true for the referee results between "2" and "5".

However, when we compared the 275 successful applications with the 275 best reviewed unsuccessful applications we found that there is hardly any difference between the average referee sores of the two groups (fig 11). Note that this group of best unsuccessful applicants is different from the one in previous sections, because the selection in this case is based on the referee scores and not on past performance.

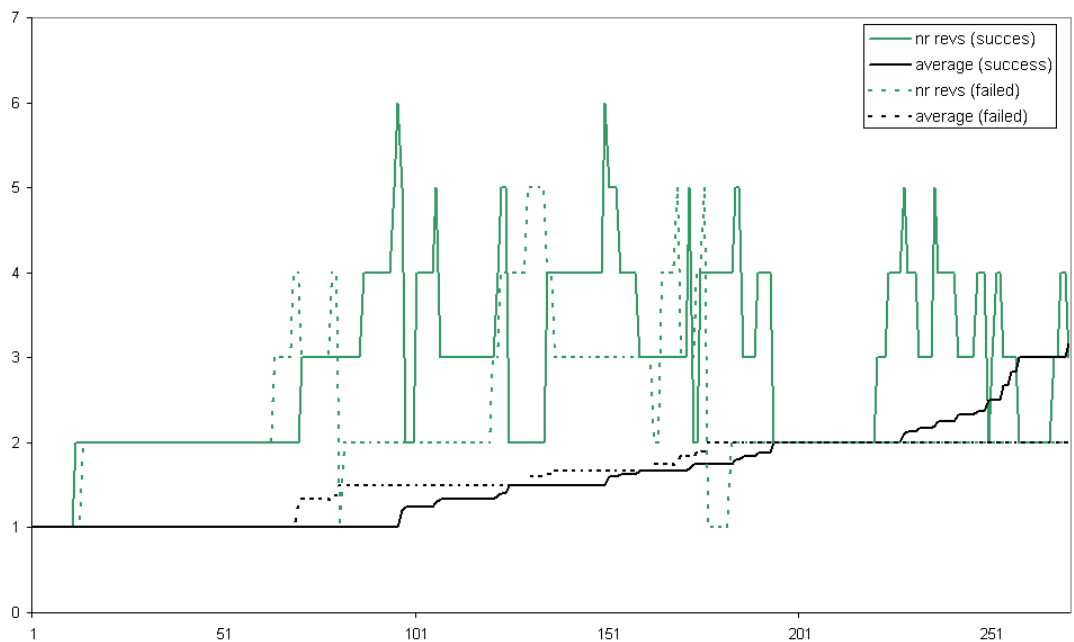

**FIGURE 11:** Average referees' score; successful applicants (black line) and the 275 best scoring unsuccessful applications (dotted black line).

A statistical analysis of the sample 275 successful and 275 best refereed unsuccessful applicants confirms this picture: ANOVA finds no differences between the two groups, no correlation exists between past performance, referee score, and funding received, and the DA does not work. In the DA, all variables are excluded from the analysis, indicating that publications, citations and referee score cannot be used to predict success.

With these results, research questions 4 and 5 can be answered:

- the funded applicants have on average a better referee's score than the non-funded applicants;
- the fundable applicants have on average a better referee's score than the non-fundable;

- the funded applicants have on average a better referee's score than the fundable-non-funded;
- but the 275 unsuccessful applicants with the highest referee scores are at the same level as the successful applicants.
- and within the group of 550 successful and best unsuccessful applicants, all relations disappear between past performance, referee score and success.

In the rest of the report, some more detailed analysis will be presented:

- Differences between disciplines: success rates, and the predictive power of past performance, network, and review results;
- If we distinguish the different instruments (Veni, Vidi, Vici, Open competition), does this influence the results?
- Gender bias?
- Do universities perform differently (and what may explain this?)
- Differences between years of application.
- What about those who not apply?

## 5.2    Disciplinary differences?

Two issues need to be addressed here. First, do the various disciplines perform differently within the total set? And secondly, are the decisions on the discipline level stronger or weaker related to past performance and the referee's judgment?

Table 15 shows two interesting patterns. First, the accepted applications tend to be unevenly distributed over the disciplines. Three disciplines have a large share of all accepted proposals: psychology and pedagogy (43%), Economics (17%), and Law (14%). All the others together only get 26%. The acceptance rates differ between the sub-disciplines, from 10% in political science to 32% in psychology.

Second, the distribution of successful applications over the disciplines reflects the size of the disciplines. The last column in table 15 gives some information about the size of the disciplines in the Netherlands universities, in terms of the number of full professors and associate professors. The differences in size are rather large. The number of professors per discipline correlates highly ($r = 0.80$) with the number of applications, and moderately high ($r = 0.53$) with the number of successful applications.

'Redistribution' takes place between the large disciplines. Economics and law get substantially less than expected given their size. Psychology gets much more: 15% of the senior staff produces 29% of the applications, and 46% of the successful one's. This is even stronger the case if we distinguish between the four instruments. The most striking point is that almost all Vici's are for researchers in the field of psychology (table 16).

**TABLE 15:** Accepted applications by subfield (OC and VI)

| | Number of applications | Applications by field | Accepted | Rejection Rate | Accepted by field | Professors* |
|---|---|---|---|---|---|---|
| Anthropology | 30 | 2.4 | 9 | 70.0 | 3.3 | 2.3 |
| Communication | 31 | 2.4 | 5 | 83.9 | 1.8 | 1.5 |
| Demography | 10 | 0.8 | 2 | 80.0 | 0.7 | 0.1 |
| Economics** | 301 | 23.7 | 46 | 84.7 | 16.8 | 35.6 |
| Education | 81 | 6.4 | 14 | 82.7 | 5.1 | 3.3 |
| Geography | 47 | 3.7 | 8 | 83.0 | 2.9 | 3.4 |
| Law | 219 | 17.2 | 41 | 81.3 | 15.0 | 28.4 |
| Political science*** | 61 | 4.8 | 6 | 90.2 | 2.2 | 6.0 |
| Psychology**** | 370 | 29.1 | 119 | 67.8 | 43.4 | 15.0 |
| Sociology | 121 | 9.5 | 24 | 80.2 | 8.8 | 4.5 |
| Total | 1271 | 100% | 274 | 78.4 | 100% | 100% |

\*  Full and associate professors by field – based on EUR, RUG, RUN, UM, UU, UvA, UvT, and VU. Source: NOD, 2004.

\*\*  Incl. management; *** Incl. public policy; **** Incl. pedagogy

**TABLE 16:** Accepted applications by subfield and instrument

| | Total | Open | | Vici | | Vidi | | Veni | |
|---|---|---|---|---|---|---|---|---|---|
| Psychology* | 119 | 65 | 41.9 | 9 | 75.0 | 18 | 42.9 | 27 | 41.5 |
| Economics** | 46 | 22 | 14.2 | 2 | 16.7 | 9 | 21.4 | 13 | 20.0 |
| Law | 41 | 31 | 20.0 | | | 4 | 9.5 | 6 | 9.2 |
| Sociology | 24 | 19 | 12.3 | | | 1 | 2.4 | 4 | 6.2 |
| Education | 14 | 2 | 1.3 | | | 2 | 4.8 | 10 | 15.4 |
| Anthropology | 9 | 5 | 3.2 | | | 3 | 7.1 | 1 | 1.5 |
| Geography | 8 | 3 | 1.9 | 1 | 8.3 | 3 | 7.1 | 1 | 1.5 |
| Political science*** | 6 | 5 | 3.2 | | | 1 | 2.4 | | |
| Communication | 5 | 2 | 1.3 | | | | | 3 | 4.6 |
| Demography | 2 | 1 | 0.6 | | | 1 | 2.4 | | |
| Total | 274 | 155 | 100% | 12 | 100% | 42 | 100% | 65 | 100% |

\* Incl. pedagogy; ** Incl. management; *** Incl. public policy

Citation behavior and publication behavior differ between disciplines and fields. Figure 12 shows the distribution of sociology and psychology journals by impact factor, and table 17 gives some statistics. Clearly, psychology journals have on average higher impact factors, and this indicates that psychologists have longer reference lists per paper than sociologist do. Also what counts as a top journal in psychology is different from what would count as top in sociology, if we take the impact factor of the journal (IF) as a criterion. The top 10% starts in sociology with an IF of 1.821, and in psychology with an IF of 3.11.

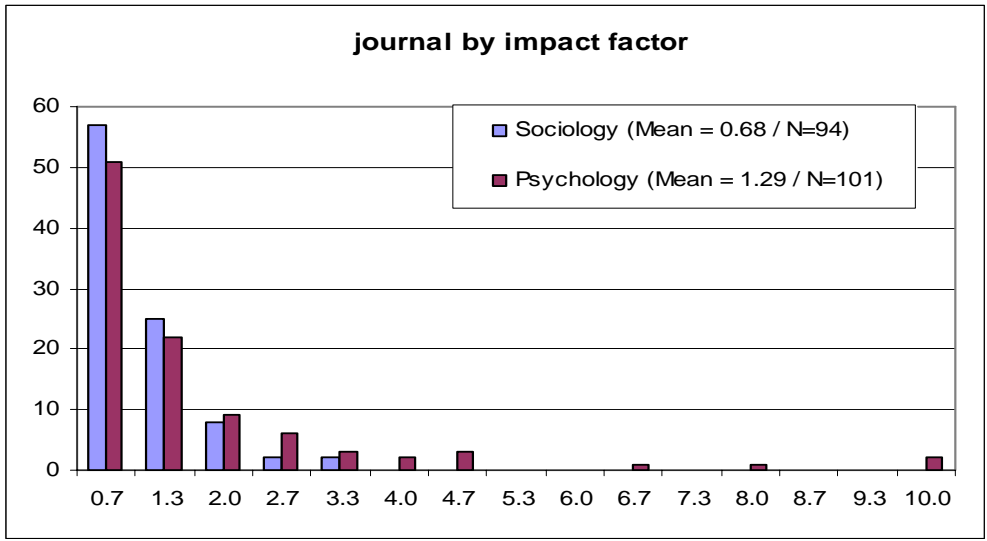

**TABLE 17:** Citation behavior by discipline

|  | mean | CoV* | median | max | top 10% | skewness | N |
|---|---|---|---|---|---|---|---|
| All SocSCI | 0.982 | 1.041 | 0.697 | 12.642 | IF >2.013 | 3.869 | 1745 |
| Psychology | 1.287 | 1.386 | 0.650 | 9.780 | IF >3.458 | 3.114 | 101 |
| Sociology | 0.683 | 0.879 | 0.460 | 3.262 | IF >1.382 | 1.982 | 94 |

*: coefficient of variance = standard deviation divided by mean; IF = Impact Factor

These differences on the discipline level are reflected in the scores on the variables used in this study. Indeed, the averages of the variables show a large variation between the disciplines (table 18).

**TABLE 18:** Past performance, network quality and referee scores by discipline

| Mean: | Pub# | cit | Pub2 | Cit2 | Pub3 | Cit3 | ref | K€ | N |
|---|---|---|---|---|---|---|---|---|---|
| Anthropology | 1.4 | 5.3 | 1.0 | 3.8 | 1.4 | 5.3 | 2.4 | 234 | 28 |
| Communication | 3.7 | 16.1 | 2.8 | 11.4 | 4.1 | 18.7 | 2.6 | 149 | 30 |
| Demography | 2.3 | 7.0 | 3.1 | 11.8 | 4.6 | 20.2 | 2.4 | 288 | 9 |
| Economics** | 2.4 | 8.5 | 2.4 | 8.8 | 3.0 | 11.4 | 2.6 | 235 | 274 |
| Education | 2.8 | 19.7 | 2.9 | 20.1 | 3.3 | 21.9 | 2.8 | 180 | 79 |
| Geography | 2.5 | 12.0 | 2.6 | 12.4 | 2.9 | 13.4 | 2.2 | 340 | 40 |
| Law | 0.5 | 2.0 | 0.6 | 2.4 | 1.0 | 3.8 | 2.3 | 187 | 206 |
| Political sci*** | 1.2 | 3.8 | 1.2 | 4.0 | 1.6 | 5.9 | 2.5 | 209 | 56 |
| Psychology* | 5.8 | 48.5 | 5.9 | 49.7 | 7.8 | 65.3 | 2.2 | 249 | 347 |
| Sociology | 2.9 | 13.3 | 3.4 | 17.3 | 4.9 | 26.2 | 2.5 | 169 | 115 |
| Total | 3.1 | 20.3 | 3.2 | 21.1 | 4.2 | 27.9 | 2.4 | 226 | 1184 |

# See table 1 for the variable names; * Incl. pedagogy; ** Incl. management; *** Incl. public policy

To find out whether these differences between disciplines influence the results of sections 5 and 6, we repeated the analysis also for the different disciplines individually. Table 19 shows the correlations between past performance, referee results and the amount of funding for the whole set, and for the disciplines individually.

**TABLE 19:** Correlations (Spearman's rho) between performance,

referees' score and received grant by discipline

|                      | PP~Rev | PP~€ | Rev~€ |    N |
|----------------------|--------|------|-------|------|
| All                  | 0.21   | 0.17 | 0.46  | 1181 |
| Average              | 0.28   | 0.28 | 0.44  |      |
| Psychology*          | 0.27   | 0.19 | 0.49  | 345  |
| Economics**          | 0.33   | 0.15 | 0.42  | 274  |
| Law                  | -      | -    | 0.43  | 206  |
| Sociology            | 0.17   | -    | 0.45  | 114  |
| Education            | 0.32   | 0.10 | 0.43  | 79   |
| Political science*** | 0.27   | -    | -     | 56   |
| Geography            | 0.36   | 0.44 | 0.37  | 40   |
| Communication        | -      | 0.28 | 0.50  | 30   |
| Anthropology         | 0.27   | 0.50 | 0.45  | 28   |
| Demography           | -      | -    | -     | 9    |

PP = past performance (average of variables pub and cit)
Netw = quality of the network (average of variables pub3 and cit3)
Rev = average of the referee scores (variable ref)
€ = amount of money received (variable euro)
* Incl. pedagogy; ** Incl. management; *** Incl. public policy

If the selected past performance indicators are valid, table 19 shows that on average these bibliometric indicators work better on the discipline level, as the correlations between these past performance indicators and the outcome of the refereeing process are slightly better here than on the more aggregated level. The same holds for the correlation between past performance and funding received. However, the correlations are still not high. The relation between the review outcome and the received funding remains unchanged. However, a few cases deviate from this pattern:

- In the case of law, the bibliometric indicators do not correlate at all with the review results and with the received funding. This is in line with the opinion that research in the field of law is not oriented at international journals but has other types of (mainly national oriented) output;
- In the case of political science, the quality indicators do not correlate with the amount of funding;
- In the case of communication, the bibliometric indicators do not correlate with the referees' evaluation;
- In the case of demography, negative correlations were found, but are not significant, due to the low *N*.

If we go to a more detailed level, the analysis generally does not improve. As an example, we calculated the correlations between past performance, the referee outcomes and the success of the application for the four sub-disciplines of psychology, as distinguished by the research council. Some of the correlation are higher than for psychology on average, others are very low and/or not significant (table 20).

**TABLE 20:** Correlations (Spearman's rho) between performance, referees' score and received grant by sub-discipline (psychology)

|  | PP~Rev | PP~€ | Rev~€ | N |
|---|---|---|---|---|
| All | 0.21 | 0.17 | 0.46 | 1181 |
| Psychology (incl ped.) | 0.27 | 0.19 | 0.49 | 345 |
| Clinical /biological /medical psychology | 0.32 |  | 0.45 | 85 |
| Developmental psychology & pedagogy | 0.25 | 0.35 | 0.49 | 68 |
| Cognitive and biological psychology | 0.28 |  | 0.49 | 109 |
| Social, work, organizational psychology; psychometrics | 0.24 | 0.29 | 0.58 | 83 |

Finally, we used Discriminant Analysis at the level of disciplines. If we only use the past performance indicators in the analysis, the results on discipline level are similar to those for the whole set, with the exception of law (83% correct positives, 15% correct negatives). But we already saw that the indicators do not seem to work in the case of law (table 19). Interestingly, if we use all the variables in a stepwise analysis, only the peer review variable (ref) is used, and the past performance indicators are removed from the analysis.

We now can answer question 6: the indicators work slightly better on discipline level, but this does not change the results obtained in the previous sections on the level of all disciplines together.[10]

# 5.3 Differences between funding instruments?

A main difference exists between the instruments: the Open Competition is, as the name suggests, open to all (teams of) senior researchers, whereas the other three programs are intended for individual researchers at different (early) stages of their career. The Veni program is for postdocs, the Vidi is for assistance of young associate professors, and the Vici is for already established but relatively young researchers who are or could become full professor.

The criteria and the selection procedure are different. In case of the open competition, the emphasis is on the quality of the proposal. In case of the Veni, Vidi, and Vici program, the individual researcher him/herself is also assessed. This would imply that in the case of an open competition, the refereeing part of the procedure is expected to be stronger than the past performance. These may be more important in the other three programs; however, the Veni program is meant for young researchers, without much past performance. Consequently, we would expect that the correlation between past performance and successful applications is lower in the open competition and in the

---

[10] If we compare the successful applicants with the best unsuccessful applicants, also on the discipline level the differences disappear. Using Anova, in each of the disciplines the referee scores, the number of publications and the citations received do not significantly differ between the succesfull and the best unsuccessful applicants.

Veni program, whereas the relation between the referee's judgment is expected to be stronger in these two cases.

Figure 21 shows the averages quality indicators by instrument. The Vici applicants are on average far more productive (publications) and visible (citations) than the others. The next highest group is the applications in the Open Completion, followed by the Vidi applicants, and the Veni applicants.

We repeated the analysis for the four programs separately, with the following results. The differences between the successful and unsuccessful applicants in the two larger programs (the Open Competition and the Veni program) are similar to the general picture of section 5 and 6. However, in case of the Vidi and Vici programs, this is not the case. The average number of publications and citations is larger in the successful group than in the unsuccessful one, but the dispersion is large (and the number of cases small), making the difference between the means not significant. However, differences in the referees' judgments are substantial and significant (table 21).

**TABLE 21:** Average quality of applicants by instrument

|              | Vici | OC   | Vidi | Veni |
|--------------|------|------|------|------|
| Publications | 9.8  | 3.8  | 3.0  | 1.7  |
| A            | 9.7  | 5.0  | 3.6  | 2.9  |
| A-           | 6.3  | 3.2  | 1.9  | 1.9  |
| B            | 10.7 | 3.7  | 3.1  | 1.4  |
| A- and B     | 9.9  | 3.4  | 2.6  | 1.5  |
| Citations    | 89   | 25   | 20   | 10   |
| A            | 112  | 39   | 25   | 21   |
| A-           | 36   | 20   | 7    | 11   |
| B            | 81   | 20   | 20   | 7    |
| A- and B     | 73   | 20   | 15   | 8    |
| Referee score| 1.70 | 2.10 | 2.30 | 2.98 |
| A            | 1.13 | 1.42 | 2.01 | 1.79 |
| A-           | 1.45 | 1.92 | 2.53 | 2.31 |
| B            | 2.06 | 2.69 | 2.51 | 2.32 |
| A- and B     | 1.75 | 2.33 | 2.52 | 3.19 |
| N            | 29   | 629  | 100  | 428  |
| A            | 12   | 155  | 43   | 65   |
| A-           | 3    | 224  | 20   | 46   |
| B            | 14   | 250  | 37   | 317  |
| A- and B     | 17   | 479  | 57   | 363  |

This is reflected in the correlation analysis (table 22). Contrary to the expectation formulated in the introduction to this section, the relationship between past performance and the referee result is moderately strong in the Vidi program, and in all the three other programs this correlation is small and non-significant. The same holds for the correlation between past performance and the research funding. The correlation between the result of the referee process and the funding awarded is the higher in the Veni program than in the open program and in Vidi. However, it is very high in case of the Vici program. Summarizing, past performance makes hardly any difference in the various programs, and in all but the Veni program the correlation is very low. This is actually unexpected, as especially in case of the Veni program for the youngest group of researchers one would not expect past performance, but promise of future performance to be dominant. Peer review is in all programs the main factor, and the correlation is especially strong in case of the Vici program.

**TABLE 22:** Correlation between quality and grant received by instrument

|              | PP~Rev | PP~€   | Rev~€ | N    |
|--------------|--------|--------|-------|------|
| All          | 0.21   | 0.17   | 0.46  | 1181 |
| Open program | 0.10   | 0.07   | 0.44  | 623  |
| Veni         | 0.31   | 0.19   | 0.49  | 426  |
| Vidi         | 0.12*  | 0.10*  | 0.41  | 100  |
| Vici         | 0.04*  | 0.17*  | 0.70  | 29   |

pp = past performance (average of variables pub and cit)
Netw = quality of the network (average of variables pub3 and cit3)
Rev = average of the referee scores (variable ref)
€ = amount of money received (variable euro)
* = non significant

The Discriminant Analysis provides us with similar results. The DA is not able to distinguish the A and A- applications. And, in case of Vidi and Vici, the DA only uses the referees' outcome and not past performance, when classifying A versus A-/B. The classification is rather good, especially in case of the Vici's: 86.2% correctly classified cases.

Again we compare the successful applications with the best scoring unsuccessful ones. Here we do this per instrument, by including the best scoring rejected applications, 154, 65, 43 and 12 for the OC, Veni, Vidi and Vici respectively. In case of the open competition, no differences exist between the successful and top-unsuccessful applications. In case of the Veni program, the successful applicants publish on average more than the unsuccessful applicants. In the other two programs, the referee score of the successful one's is on average better than the scores of the unsuccessful applications. The correlation analysis shows the same, as table 23 shows.

**TABLE 23:** Correlation between quality and grant received by instrument (550 cases)

|              | PP~Rev | PP~euro | Rev~euro | N   |
|--------------|--------|---------|----------|-----|
| All          | 0.03*  | 0.07*   | 0.07*    | 550 |
| Open program | - 0.09* | - 0.01* | 0.00*    | 304 |
| Veni         | 0.22   | 0.13*   | 0.05*    | 132 |
| Vidi         | .05*   | 0.10*   | 0.31     | 88  |
| Vici         | 0.06*  | 0.17*   | 0.65     | 24  |

pp = past performance (average of variables pub and cit)
Rev = average of the referee scores (variable ref)
euro = amount of money received (variable euro)
* = non significant

The results of the Discriminant Analysis are similar[11]. In case of the OC, the classification is worse than a random classification; in case of the Veni program, the about half of the successful and half of the unsuccessful applications are classified correctly. As the prior probabilities in this sample are also 50%, the past performance variables and the referee variable do not improve the classification. In the two other programs (Vidi and Vici), especially the referee variable leads to a correct classification of about 70 to 80 % of the cases.

If we draw a conclusion on the level of the individual instruments, success is not or hardly related to past performance (table 22 and 23). Success correlates with referee judgments, but if we restrict the analysis to the best 550 applicants, this is not the case anymore in the OC and the Veni program. Finally, on the level of the individual programs, reviews and past performance hardly correlate, and only for the Veni program correlation between past performance and review score is moderately strong.

**TABLE 23:** Classification by instrument (550 cases)

| Instrument | | | Funded | Predicted Group Membership | | Total |
|---|---|---|---|---|---|---|
| | | | | 1 | 2 | |
| OC* | Original | Count | Yes | 61 | 92 | 153 |
| | | | No | 78 | 73 | 151 |
| | | % | Yes | 39.9 | 60.1 | 100.0 |
| | | | No | 51.7 | 48.3 | 100.0 |
| Veni** | Original | Count | Yes | 30 | 37 | 67 |
| | | | No | 27 | 37 | 64 |
| | | % | Yes | 44.8 | 55.2 | 100.0 |
| | | | No | 42.2 | 57.8 | 100.0 |
| Vidi*** | Original | Count | Yes | 28 | 16 | 44 |
| | | | No | 12 | 31 | 43 |
| | | % | Yes | 63.6 | 36.4 | 100.0 |
| | | | No | 27.9 | 72.1 | 100.0 |
| Vici**** | Original | Count | Yes | 10 | 2 | 12 |
| | | | No | 2 | 10 | 12 |
| | | % | Yes | 83.3 | 16.7 | 100.0 |
| | | | No | 16.7 | 83.3 | 100.0 |

*     44.1% of original grouped cases correctly classified.
**    51.1% of original grouped cases correctly classified.
***   67.8% of original grouped cases correctly classified.
****  83.3% of original grouped cases correctly classified.

# 5.4     Gender differences

In the available dataset, 32% of all applications have a female principle investigator, unevenly distributed over the  disciplines. Relatively more female than male applicants are found in anthropology, psychology and education. The opposite is the case in economics (table 25).

---

[11] The stepwise procedure does not work in case of the OC.

**FIGURE 25:** Gender of applicants by discipline

|  | Male | | Female | | Total | |
|---|---|---|---|---|---|---|
| Anthropology | 11 | 1.4% | 17 | 4.5% | 28 | 2.4% |
| Communication | 18 | 2.2% | 12 | 3.2% | 30 | 2.5% |
| Demography | 4 | .5% | 5 | 1.3% | 9 | .8% |
| Economics** | 225 | 27.8% | 49 | 13.0% | 274 | 23.1% |
| Education | 30 | 3.7% | 49 | 13.0% | 79 | 6.7% |
| Geography | 35 | 4.3% | 5 | 1.3% | 40 | 3.4% |
| Law | 144 | 17.8% | 62 | 16.4% | 206 | 17.4% |
| Political sci*** | 44 | 5.4% | 12 | 3.2% | 56 | 4.7% |
| Psychology* | 219 | 27.1% | 128 | 33.9% | 347 | 29.3% |
| Sociology | 78 | 9.7% | 39 | 10.3% | 117 | 9.9% |
| Total | 808 | 100% | 378 | 100% | 1186 | 100% |

* Incl. pedagogy; ** Incl. management; *** Incl. public policy

Of the funded applications, 29% have a female applicant. Some 27% of the A-applications are from female researchers, and a little more than 35% of the B applications. As far as successful applications are concerned, women are as successful as men in law and psychology, but much less in economics, anthropology, communication, sociology and political science (table 26). Here, we can only speculate why this is the case.

We also find differences between the funding instruments and between disciplines. Women are more successful than men in the open competition and in the Vidi program, but less successful in the Veni and Vici programs.

**TABLE 26:** Success by gender and discipline

|  |  | male | Female |  | male | Female |
|---|---|---|---|---|---|---|
| Anthropology | A | 55% | 17% | Geography | 20% | 20% |
|  | A- | 9% | 35% |  | 31% | 20% |
|  | B | 36% | 47% |  | 49% | 60% |
|  | N | 11 | 17 |  | 35 | 5 |
| Communication | A | 22% | 8% | Law | 20% | 19% |
|  | A- | 22% | 17% |  | 30% | 39% |
|  | B | 56% | 75% |  | 50% | 42% |
|  | N | 18 | 12 |  | 144 | 62 |
| Demography | A | 25% | 20% | Political science | 11% | 8% |
|  | A- | 25% | 60% |  | 23% | 17% |
|  | B | 50% | 20% |  | 66% | 75% |
|  | N | 4 | 5 |  | 44 | 12 |
| Economics | A | 19% | 8% | Psychology | 34% | 34% |
|  | A- | 26% | 14% |  | 26% | 16% |
|  | B | 56% | 78% |  | 40% | 49% |
|  | N | 225 | 49 |  | 219 | 128 |
| Education | A | 20% | 16% | Sociology | 24% | 15% |
|  | A- | 20% | 8% |  | 32% | 21% |
|  | B | 60% | 76% |  | 44% | 64% |
|  | N | 30 | 49 |  | 78 | 39 |

As gender inequality is increasingly becoming a policy issue, are differences between men and women changing?  We do not have a long time series, but only three years of observation (table 27).Real trends are not particularly observable. Nevertheless, in 2005 the picture differs from the earlier years.  In 2005, some 60% of as well the male as the female applicants is in the B category. And a larger percentage of women (24%) is successful in 2005 than male researchers (20%).

**TABLE 27:** Success by gender

|        |      | Male | Female | Total |
|--------|------|------|--------|-------|
|        | 2003 | 26   | 21     | 25    |
| A      | 2004 | 25   | 19     | 23    |
|        | 2005 | 20   | 24     | 21    |
|        | 2003 | 24   | 15     | 21    |
| A-     | 2004 | 36   | 30     | 34    |
|        | 2005 | 19   | 15     | 18    |
|        | 2003 | 50   | 36     | 46    |
| A + A- | 2004 | 61   | 49     | 57    |
|        | 2005 | 40   | 39     | 39    |
|        | 2003 | 50   | 64     | 54    |
| B      | 2004 | 39   | 51     | 43    |
|        | 2005 | 60   | 61     | 61    |
|        | 2003 | 272  | 99     | 371   |
| Total  | 2004 | 288  | 144    | 432   |
|        | 2005 | 248  | 135    | 383   |

## Gender bias?

The question of gender bias in (peer) review procedures is an important issue (Wenneras & Wold 1997). Empirical research shows that there are contradictory findings, although a recent meta-analysis suggests that (small) gender bias does exist (Bornmann et al, forthcoming).

Here, we define gender inequality as men having a better chance to get a project funded than women with the same referee result and with the same past performance.  For example, if male researchers have a better past performance, this may explain the higher rating by the referee's and it may explain a higher success rate of male researchers. Table 25 shows that male and female researchers indeed do differ significantly in terms of publications, citations, and referee results. Male researchers get higher referee scores. The average score of female researchers is 87% of the score of male researchers (table 28). Female researchers also have lower past performance (about two third) than male researchers have.

If we distinguish between three groups of A, A- and B applications, the picture becomes slightly different. In all cases, male researchers score better. But for the A and A- applications, the difference between the number of citations for male and for female researchers' is not any more significant. As table 28 shows, male and female researchers actually receive about the same number of citations in the A and A- group. The differences between the number of publications and between the refereeing results remain significant (at 10%). In the B category, differences between male and female researchers are significant.

**TABLE 28:** Applications by gender

|  | Average nr of publications | Average nr of citations | Average referee score | N |
|---|---|---|---|---|
| Male | 3.52 | 23 | 2.32 | 808 (68.1%) |
| Female | 2.24 | 16 | 2.66 | 378 (31.9%) |
| Female / male | .64 | .69 | .87 | |
| | | | | |
| Male A | 4.70 | 37 | 1.55 | 194 (70.5%) |
| Female A | 3.84 | 34 | 1.71 | 81 (29.5%) |
| Female / male | .82 | .92 | .91 | |
| | | | | |
| Male A- | 3.24 | 18 | 1.98 | 215 (73.4%) |
| Female A- | 1.99 | 16 | 2.15 | 78 (26.6%) |
| Female / male | .58 | .89 | .92 | |
| | | | | |
| Male B | 3.10 | 18 | 2.87 | 399 (64.6%) |
| Female B | 1.73 | 9 | 3.20 | 219 (35.4%) |
| Female / male | .56 | .50 | .90 | |

Does the selection process of the research council shows a gender bias?
- First, based on the findings in table 28, female researchers are not disadvantaged in the *refereeing process* because the gender difference is smaller than it is in past performance.[12]
- Second, cross tabs show a significant correlation between gender and the decision about the application (A, A-, B). Significantly more applications by women get the verdict 'not fundable' than applications by men (49% for men versus 58% for women). However, within the set of fundable (A & A-) projects, 47% of the men and 51% if the women get funded. This difference is not statistically significant. So in the higher categories, no significant gender difference is found.
- Finally, we calculated the share of female researchers in the 275 best reviewed applications, in the set of 275 most publishing applicants and in the set of 275 most cited applicants (table 29). In all the three 'top lists' the share of female researchers is smaller than the share of women in the set of successful applicants.

---

[12] The correlations between past performance and referee result are low but similar for men and women. And, the relative (to men) referee results for women are not lower than the relative past performance scores (table 27). This suggests that women do not get lower referee scores than men with similar past performance.

**TABLE 29:** Gender bias?

| | Male | Female | % Female |
|---|---|---|---|
| Successful 275 applicants | 194 | 81 | 29.5% |
| Top 275 refereed applications | 211 | 64 | 23.3% |
| Top 275 publishing applicants | 218 | 57 | 20.7% |
| Top 275 cited applicants | 211 | 64 | 23.3% |

Overall, this shows that the final decision-making corrects in favor of female applicants – reflecting a policy of stimulating women to go for research careers.

## 5.5 Differences between universities

In this section we give only the distribution of successful applications by university and by instrument.  Here, the differences are considerable. Of course, the differences may reflect size of the social, behavioral, economics and law faculties in the various universities. It also may reflect the different sizes of the subfields within the universities: a large psychology faculty is helpful as almost 50% of the successful applications is in psychology.

**TABLE 30:** Successful applications by instrument and university

| | Total | | OC | | Vici | | Vidi | | Veni | |
|---|---|---|---|---|---|---|---|---|---|---|
| UvA | 42 | 15.3% | 18 | 11.6% | 4 | 33.3% | 6 | 14.0% | 14 | 21.5% |
| UvT | 40 | 14.5% | 26 | 16.8% | 1 | 8.3% | 6 | 14.0% | 7 | 10.8% |
| RUN | 33 | 12.0% | 20 | 12.9% | 1 | 8.3% | 6 | 14.0% | 6 | 9.2% |
| UU | 31 | 11.3% | 15 | 9.7% | 2 | 16.7% | 2 | 4.7% | 12 | 18.5% |
| UM | 26 | 9.5% | 17 | 11.0% | 2 | 16.7% | 2 | 4.7% | 5 | 7.7% |
| VU | 25 | 9.1% | 13 | 8.4% | 0 | .0% | 5 | 11.6% | 7 | 10.8% |
| UL | 22 | 8.0% | 14 | 9.0% | 0 | .0% | 4 | 9.3% | 4 | 6.2% |
| RUG | 21 | 7.6% | 13 | 8.4% | 0 | .0% | 6 | 14.0% | 2 | 3.1% |
| EUR | 9 | 3.3% | 4 | 2.6% | 0 | .0% | 1 | 2.3% | 4 | 6.2% |
| WUR | 7 | 2.5% | 5 | 3.2% | 0 | .0% | 2 | 4.7% | 0 | .0% |
| TUD | 2 | .7% | 1 | .6% | 0 | .0% | 1 | 2.3% | 0 | .0% |
| TUE | 2 | .7% | 2 | 1.3% | 0 | .0% | 9 | .0% | 0 | .0% |
| UT | 3 | 1.1% | 2 | 1.3% | 0 | .0% | 0 | .0% | 1 | 1.5% |
| Other | 12 | 4.4% | 5 | 3.2% | 2 | 16.7% | 2 | 4.7% | 3 | 4.6% |
| | 275 | 100% | 155 | 100% | 12 | 100% | 43 | 100% | 65 | 100% |

## 5.6 Network effects: the co-applicants

Another issue we explore here is the effect of the network of the main applicant. As argued above, we take co-applicants as a proxy for the network. Only open competition applications are included in the analysis, as the VI programs almost always have only one applicant.

Within the open competition applications, 51% of the applications have two applicants, and 21% of the applications have three or more. This means that only 28% of the applications have only one applicant. Does the past performance of the co-applicants play a role in the success of an application? The results of the analysis are as follows:

- The network quality as defined is section 3 is significantly better for the A than for the A- applications, and the latter are again better than the B applications.
- Correlation between the network indicators and the referee outcome is low to moderate, but higher than the correlation between the past performance and the result of the refereeing process.
- The same holds for the correlation between network indicators and the amount of funding.
- Including the network indicators in the Discriminant Analysis hardly improves the classification.

We may therefore conclude that the quality of the co-applicants does positively influence the success of an application, but the effect is not very strong.

**TABLE 31:** Network effects (Spearman's rho)

|  | average score referees |  | money received |  |
|---|---|---|---|---|
| Publications main applicant | .077 |  | .103 | (**) |
| Network: |  |  |  |  |
|   Average publications all (co-)applicants | .089 | (*) | .143 | (**) |
|   Publications most productive (co-)applicant | .076 |  | .149 | (**) |
|  |  |  |  |  |
| Citations main applicant | .113 | (**) | .133 | (**) |
| Network: |  |  |  |  |
|   Average nr citations all (co-)applicants | .127 | (**) | .165 | (**) |
|   Citations most visible (co-)applicant | .124 | (**) | .168 | (**) |

\* Correlation is significant at the 0.05 level (2-tailed).
\*\* Correlation is significant at the 0.01 level (2-tailed).
N = 623 (open competition only)

# 5.7 Differences between accepted, rejected and non-applicants (vintage 2003)

As the council's goal is to fund better researchers, one would expect that in the clientele of the council the better researchers are over-represented. In this section, we test whether the applicants differ from the non-applicants in terms of past performance.

We ran an ANOVA to test statistically whether the recipients score higher in terms of output and citations than the researchers that did not get their proposal accepted, and researchers who did not apply. This analysis is done for 2003 only, as an initial test which includes the non-applicants.

The three groups score differently in terms of the average numbers of publications and received citations (table 32). The successful applicants publish more than the failed applicants and the non-applicants) (4.37 versus 2.87 versus 1.99 publications) and are more cited (47 versus 21 versus 20 citations). A post hoc test (Scheffe) shows that the differences between group 1 and 2, and between group 1 and 3 are statistically significant, but the difference between group 2 and 3 is not.

**TABLE 32:** Average number of publications and citations by group

| | | N | Mean | Std. Dev | Std. Error | 95% Confidence Interval for Mean Lower Bound | 95% Confidence Interval for Mean Upper Bound | Minimum | Maximum |
|---|---|---|---|---|---|---|---|---|---|
| PUB | 1 | 93 | 4.37 | 5.628 | .584 | 3.21 | 5.52 | 0 | 29 |
| | 2 | 302 | 2.87 | 5.389 | .310 | 2.26 | 3.48 | 0 | 62 |
| | 3 | 11761 | 1.99 | 2.663 | .025 | 1.95 | 2.04 | 1 | 65 |
| | Total | 12156 | 2.03 | 2.807 | .025 | 1.98 | 2.08 | 0 | 65 |
| CIT | 1 | 93 | 47.27 | 84.767 | 8.790 | 29.81 | 64.73 | 0 | 593 |
| | 2 | 302 | 20.97 | 54.560 | 3.140 | 14.80 | 27.15 | 0 | 524 |
| | 3 | 11761 | 20.14 | 45.891 | .423 | 19.31 | 20.97 | 0 | 1115 |
| | Total | 12156 | 20.37 | 46.596 | .423 | 19.54 | 21.20 | 0 | 1115 |

| | | Sum of Squares | df | Mean Square | F | Sig. |
|---|---|---|---|---|---|---|
| Pub | Between Groups | 735.692 | 2 | 367.846 | 47.034 | .000 |
| | Within Groups | 95047.140 | 12153 | 7.821 | | |
| | Total | 95782.832 | 12155 | | | |
| Cit | Between Groups | 68020.805 | 2 | 34010.402 | 15.702 | .000 |
| | Within Groups | 26323315.863 | 12153 | 2165.993 | | |
| | Total | 26391336.668 | 12155 | | | |

(1 = successful applicants; 2 = failed applicants; 3 = non applicants)

Did the researchers with a higher performance also receive more funding in 2003? To answer this question, we correlated the variables 'publications' and 'citations' with the variable 'euro received'. Publications are correlated marginally more than citations with obtaining grants, but the Pearson correlation coefficients are barely larger than zero.[13] All correlations are significant because of the large N. The high correlation between publications and citations was expected.

**TABLE 33:** Pearson correlation between publications, citations, and received funding – including non-applicants

| | | Cit | euro |
|---|---|---|---|
| Pub | Pearson Correlation | .730(**) | .071(**) |
| | Sig. (2-tailed) | .000 | .000 |
| | N | 12156 | 12156 |
| Cit | Pearson Correlation | | .049(**) |
| | Sig. (2-tailed) | | .000 |
| | N | | 12156 |

** Correlation is significant at the 0.01 level (2-tailed).

13 Excluding the applicants without ISI publications from the analysis does not influence the results.

Again, the distribution of the data (citations, publications) is rather skewed; therefore a rank order correlation is preferred. If we use a correlation measure for ordinal data (Spearman's rho), the correlations between publications and citations with grants received is even smaller.

**TABLE 34:** Correlation (rho) between publications, citations, and received funding – including non-applicants

| | | Cit | euro |
|---|---|---|---|
| Pub | Spearman's rho | .541(**) | .032(**) |
| | Sig. (2-tailed) | .000 | .000 |
| | N | 12156 | 12156 |
| Cit | Spearman's rho | | .017 |
| | Sig. (2-tailed) | | .064 |
| | N | | 12156 |

** Correlation is significant at the 0.01 level (2-tailed).

If one accepts the definition of quality in terms of these indicators, then an interesting question is the extent to which the quality of the researcher predicts whether or not he/she gets a proposal accepted. Using Discriminant Analysis, we used the variables pub and cit as predictors for group membership: success, failure, or non-applicant. Only in one third of the cases, could the successful applicants be predicted, and only in 20% the rejected applications. The correct predictions are overshadowed by the size of Group 3 (the non-applicants).

In summary, applicants perform on average better than the others (i.e., rejected and non-applicants). But the predictive value of past performance is rather low, and correlation between past performance and receiving research money is low. If we restrict ourselves to the top 275 non applicants, as we have done in the other analyses shown in this report, this may again give an opposite result.

**TABLE 35:** Classification results – three groups (a)

| | | Predicted Group Membership | | | Total |
|---|---|---|---|---|---|
| | Group (b) | 1 | 2 | 3 | |
| Count | 1 | 31 | 19 | 43 | 93 |
| | 2 | 62 | 59 | 181 | 302 |
| | 3 | 1016 | 1820 | 8925 | 11761 |
| % | 1 | 33.3 | 20.4 | 46.2 | 100.0 |
| | 2 | 20.5 | 19.5 | 59.9 | 100.0 |
| | 3 | 8.6 | 15.5 | 75.9 | 100.0 |

a. 72.2% of original grouped cases correctly classified.
b. 1 = successful applicants; 2 = failed applicants; 3 = non-applicants

**TABLE 36:** Classification results – Wilks' Lambda

| Test of Function(s) | Wilks' Lambda | Chi-square | Df | Sig. |
|---|---|---|---|---|
| 1 through 2 | .990 | 121.862 | 4 | .000 |
| 2 | .999 | 15.251 | 1 | .000 |

# 6    Conclusion and discussion

## 6.1    Conclusions

The selected performance indicators:

- ▪ Publications can be considered an indicator of productivity; citations as an indicator of diffusion. In this study, citations ("visibility", "diffusion") were shown to be a somewhat better indicator than publications ("productivity") for predicting success in the case of a grant application. Differentiation among the disciplines did not change the results of the analysis;

- ▪ The successful applicants are not so much more productive, but on average, are more visible (cited) than the failed applicants. Stepwise analysis removes the variable "publications" from the Discriminant Analysis (DA);

- ▪ The number of "false positives" is (much) larger than the number of "false negatives". This means that the system is more successful in declining applications than awarding them. At most, one third of the awards are positively indicated by these measures of past performance.

- ▪ The definition of past performance takes into account only a part of the research output. E.g., books and book chapters are not included, although they are important in some of the disciplines considered. However, the conclusions remain valid when the analysis is restricted to disciplines where international journals are the dominant form of output.

Peer review:

- ▪ The outcome of the peer review is a stronger predictor of application success. It correlates low with past performance indicators, but much higher with the funding received. However the correlation is less than 0.5, which leaves much of the variance unexplained. In other words, the discretional power of the council is large. Peer review can be considered as part of the external organization of the council;

- ▪ In the classification of successful and unsuccessful applications, the peer review result remains in the analysis with the citations received, whereas the number of publications is removed from the stepwise analysis. About 85% of the positives and about 65% of the negatives are in this case correctly classified;

- ▪ We found a large variation in number of reviews, and the approved applications have on average more reviews than those that were not-approved: "A applications" got on average 2.9 reviews, 'A- applications' on average 2.6 and 'B applications' got on average 2.2 reviews. This raises questions with respect to the procedures; however these differences are due in part to the preselection phase in some of the programs.

Effect of the skewed distributions

- The results change if we compare successful applications with a set of a similar size, consisting of the best of the rejected applications. Differences in past performance and in referee's judgments disappear, or the relationships change their sign. Review scores and past performance are in this case negatively related to funding;
- This suggests that equally good applications are currently not funded, and may be considered as a loss of talent.

Differences between MaGW programs?

- Differences were found between MAGW programs. On the level of the individual programs similar results were observed as those found for the whole set.
- However, the picture changes here if we only include the best rejected applications. The correlation between past performance and success disappears, but referee scores are correlated moderately to strongly with success. The relation between referee outcome and past performance also completely disappears;
- In the OC, the quality of the proposal is expected to be the central criterion, and in the VI, the quality of the applicant. Our analysis does not support this.

Differences between disciplines?

- First, the distribution of funds over disciplines is proportional to the size of senior staff in various disciplines across the universities. The share of psychology (incl. pedagogy) in the awarded grants is much larger than one would expect on the basis of the size of the field. For economics and law, the opposite holds;
- Second, at the discipline level the indicators work slightly better and the correlations are higher. This was expected. Peer review based indicators are more adequate on a low level of aggregation, and going from discipline to sub-discipline only slightly improves the results.

Gender differences?

- Female researchers receive lower grades from the referees;
- Female researchers also score lower on past performance indicators;
- If we compare male and female researchers in the higher categories (A & A-), women still score lower than men on all variables, but the differences are relatively small and not significant;
- Gender differences vary across the disciplines, and also across different instruments;
- Over time, female researchers are becoming relatively more successful, and even relatively more successful than men. In 2005, 24% of the applications of women were successful, against 20% of male researchers;
- Decision-making by the council is favorable for female researchers.

Effect of quality of the network?

▪ The past performance of the co-applicants is positively related to success. The effect is not very strong.

## 6.2 Open questions and policy implications

This research leads to several open questions. Firstly, correlations between past performance and referee score's are relatively low, so much of the variance is unexplained. A low correlation between academic quality and the peer review outcome indicators, on the one side, and research funding, on the other, suggest that these factors are not used to support the decision-making processes. Further research on what happens during the *decision-making process* is needed, due to the fact that a (high) percentage of the variance remains unexplained. Other factors and criteria seem to dominate the decision-making process. Candidates for this may be the following:
- the assessment (by the decision-making council) of which proposals are better or more doable, which ones focus on more relevant topics or fields, or which ones provide the promise of more scientific progress.
- the discipline: the distribution of funds varies enormously among disciplines. Does this reflect quality? Is psychological research better than the research carried out in other fields?

The *peer review* of the proposals is much more influential than the (also *peer review* based) bibliometric indicators, because it correlates much better with the funding received. Evaluation of the peer review process is therefore needed. The claim that distributing research grants through NWO leads to the selection of the best researchers needs further in-depth evaluation. Who are the reviewers, and how are they selected? And, could the nature of the network relations between applicants, co-applicants, reviewers and decision-makers explain success and failure? Comparison with other systems (NSF, DFG, for example) may be instructive.

Other issues for further research:
- What about *post performance*? Do researchers funded by MaGW perform better ex post?
- This study covers only one research council. Extension to others would be useful.

The following *policy issues* can be derived from this study:
- The analysis shows that the difference between poor and good is easier to define, than it is between good and very good. This suggests that certain procedural changes may be useful. Perhaps only a first round of rejecting poor applications is needed, and then a light selection of the winners. Perhaps it is also useful to make a random selection from the set of good applications/applicants.

- Indicators work slightly better at the *disciplinary level* than at the level of the social sciences in general. Since a quality comparison cannot be made on the higher MAGW level, does this imply that one should return to a more decentralized allocation mechanisms?

On a more fundamental level the question emerges of whether the project selection mechanism is based on an unwarranted rationalist model: is it possible to pick the winners at the individual level?  The criteria and indicators are however, never unambiguous, and this holds for bibliometric indicators as well as for peer review.

Perhaps the quality of project allocation is more a systems level issue. Instead of focusing on processes for selecting individual projects, we may have to ensure that the system works properly:

- Quality requires variation and selection
- Is variation supported? (by different funding institutions)
- Is selection adequate? (a variety of criteria, open for innovation)
- Are roles assigned adequately?
- Is the system evaluated regularly?

## Recent *Science System Assessment* reports: